%% file: grf.tex
\documentclass[sigconf]{acmart}

\usepackage{xcolor}
\usepackage{soul}
\acmSubmissionID{947}

\AtBeginDocument{%
  \providecommand\BibTeX{{%
    \normalfont B\kern-0.5em{\scshape i\kern-0.25em b}\kern-0.8em\TeX}}}

\copyrightyear{2023}
\acmYear{2023}
\setcopyright{acmlicensed}\acmConference[SA Conference Papers '23]{SIGGRAPH Asia 2023 Conference Papers}{December 12--15, 2023}{Sydney, NSW, Australia}
\acmBooktitle{SIGGRAPH Asia 2023 Conference Papers (SA Conference Papers '23), December 12--15, 2023, Sydney, NSW, Australia}
\acmPrice{15.00}
\acmDOI{10.1145/3610548.3618247}
\acmISBN{979-8-4007-0315-7/23/12}

\begin{document}

\title{GroundLink: A Dataset Unifying Human Body Movement and Ground Reaction Dynamics}

\author{Xingjian Han}
\orcid{0000-0002-1899-1952}
\affiliation{%
 \department{Computer Science}
  \institution{Boston University}
  \city{Boston}
  \country{USA}
}
\email{xjhan@bu.edu}

\author{Benjamin Senderling}
\orcid{0000-0003-2502-0553}
\affiliation{%
 \department{Physical Therapy}
  \institution{Boston University}
  \city{Boston}
  \country{USA}
}
\email{bsender@bu.edu}

\author{Stanley To}
\orcid{0009-0000-8702-0124}
\affiliation{%
 \department{Computer Science}
  \institution{Boston University}
  \city{Boston}
  \country{USA}
}
\email{stanleyto275@gmail.com}

\author{Deepak Kumar}
\orcid{0000-0003-1728-484X}
\affiliation{%
 \department{Physical Therapy}
  \institution{Boston University}
  \city{Boston}
  \country{USA}
}
\email{kumard@bu.edu}

\author{Emily Whiting}
\orcid{0000-0001-7997-1675}
\affiliation{%
 \department{Computer Science}
  \institution{Boston University}
  \city{Boston}
  \country{USA}
}
\email{whiting@bu.edu}

\author{Jun Saito}
\orcid{0000-0002-3128-9009}
\affiliation{%
  \institution{Adobe Research}
  \city{Seattle}
  \country{USA}
}
\email{jsaito@adobe.com}


\begin{abstract}
The physical plausibility of human motions is vital to various applications in fields including but not limited to graphics, animation, robotics, vision, biomechanics, and sports science. While fully simulating human motions with physics is an extreme challenge, we hypothesize that we can treat this complexity as a black box in a data-driven manner if we focus on the ground contact, and have sufficient observations of physics and human activities in the real world. To prove our hypothesis, we present \textit{GroundLink}, a unified dataset comprised of captured ground reaction force (GRF) and center of pressure (CoP) synchronized to standard kinematic motion captures. GRF and CoP of \textit{GroundLink} are not simulated but captured at high temporal resolution using force platforms embedded in the ground for uncompromising measurement accuracy. This dataset contains 368 processed motion trials ($\sim 1.59M$ recorded frames) with 19 different movements including locomotion and weight-shifting actions such as tennis swings to signify the importance of capturing physics paired with kinematics. \textit{GroundLinkNet}, our benchmark neural network model trained with \textit{GroundLink}, supports our hypothesis by predicting GRFs and CoPs accurately and plausibly on unseen motions from various sources. The dataset, code, and benchmark models are made public for further research on various downstream tasks leveraging the rich physics information at \href{https://csr.bu.edu/groundlink/}{https://csr.bu.edu/groundlink/}. 
\end{abstract}


\begin{teaserfigure}
  \includegraphics[width=\textwidth]{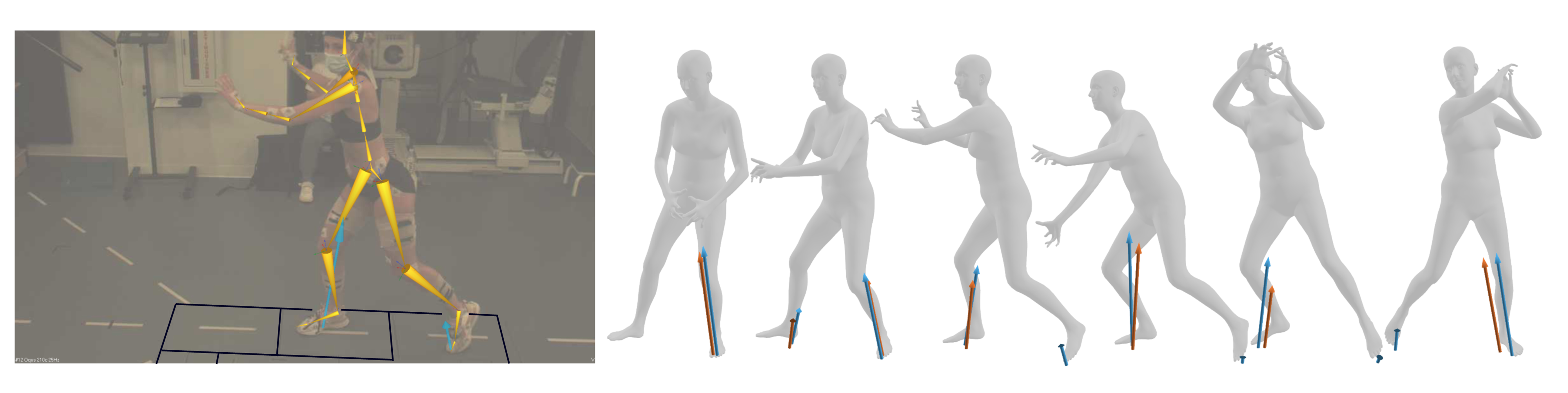}
  \caption{
  \textit{GroundLink} is a dataset unifying standard kinematic motion capture with real-world physics information including ground reaction force (GRF) and center of pressure (CoP). (Left) The data is captured and synchronized at high temporal resolution in a biomechanics laboratory with mechanical force plates embedded in the ground and multiple high-speed cameras for uncompromising measurement accuracy. Our benchmark model \textit{GroundLinkNet} trained with this data demonstrates its generalization capability by predicting GRFs and CoPs on unseen motions accurately and plausibly. (Right) Ground truth reaction forces (blue) compared to forces predicted by \textit{GroundLinkNet} (orange) for a sample tennis motion.
  }
  \label{fig:teaser}
\end{teaserfigure}

\maketitle

\input{SecIntro}
\input{SecMocap}
\input{SecNN}
\input{SecResult.tex}

\begin{acks}
    The authors would like to thank the anonymous reviewers for their valuable comments. The authors also thank the participants for their time and commitment to the study. This project was partially supported by the Alfred P. Sloan Foundation: Sloan Research Fellowship.
\end{acks}
%
\bibliographystyle{ACM-Reference-Format}
\bibliography{grf}

\clearpage
\input{SecFigsOnly.tex}

\end{document}

%% file: SecIntro.tex
\section{Introduction}
\begin{figure*}[t]
\centering
\setlength{\unitlength}{\linewidth}
  \begin{picture}(1,0.25)%
      \put(0,0){\includegraphics[width=\linewidth]{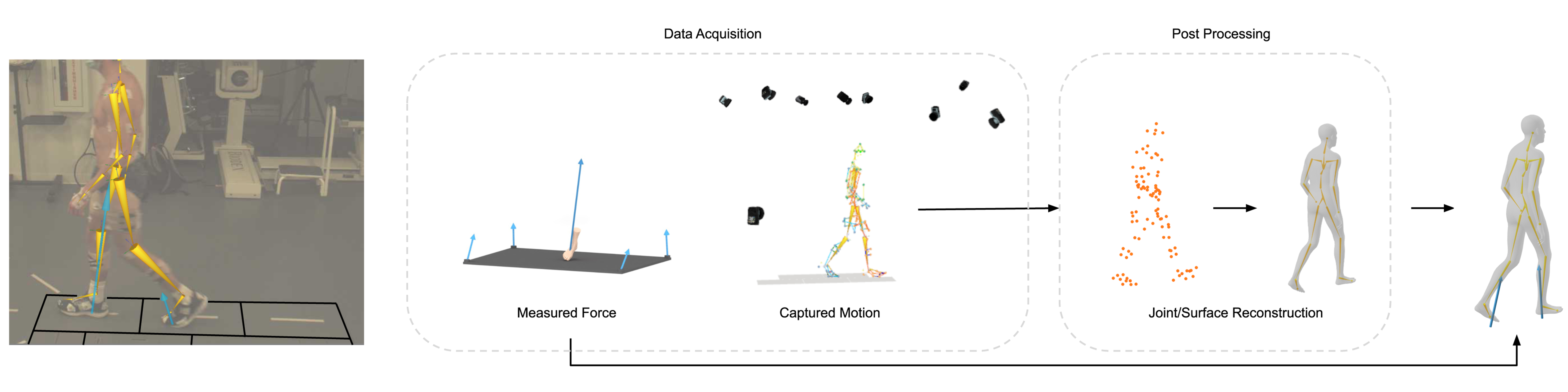}}%

    \small
  \end{picture}
\vspace{-20pt}
\caption{
\textit{GroundLink} pipeline. The Data Acquisition phase uses mechanical force platforms with high sampling frequency and an infrared motion capture system (Qualisys). Center of pressure (CoP) is obtained from the ground reaction force and moments (GRF\&M). In the Post Processing phase, we annotate the MoCap raw marker data and optimize the joint angles and shape to obtain the input motion sequence parameters. Together, the resulting dataset consists of force data (GRF and CoP) synchronized with the corresponding MoCap sequences. 
}
\label{fig:pipeline}
\end{figure*}

Accurate and realistic representations of the human body and dynamics are important for graphics applications such as animation production and biomedical applications such as clinical movement analysis or medical device design. Understanding the underlying physics quantities and kinematics parameters, such as ground reaction force and moment, center of pressure, joint torque, and body trajectories, is critical for analyzing human movement dynamics.

In contrast to the growing demand for convincing human motions, most existing motion capture datasets focus on kinematics (e.g.,~\cite{cmu, accad, taichiScott2020}) without physics quantities associated with them.

To address the need, we introduce \textit{GroundLink}, a public dataset that consists of motion capture data and synchronized ground reaction forces (GRF) and center of pressure (CoP) captured from laboratory force plates. To fill the gap in the current demand for physically-plausible human movement the dataset includes subtle and slow movements such as yoga, physical therapy exercises, tennis swing motions, and walking. For example, Fig.~\ref{fig:teaser} shows a noticeable weight transfer,
indicating an intentional adjustment to maintain balance. This is the first public motion dataset in the graphics community that consists of laboratory-based ground reaction force measurements for various human motions. 

To demonstrate the effectiveness of the kinematics data paired with physics quantities, we conduct various experiments with our benchmark deep neural network model \textit{GroundLinkNet} trained with \textit{GroundLink}. While its architecture is straightforward, we show that the model generalizes well in predicting GRFs and CoPs given the unseen kinematic motion sequences and body shapes. We believe this is a promising sign that a reasonable amount of kinematics data coupled with real-world physics quantities enable us to build generalizable human motion models with the understanding of physics for numerous applications in graphics, vision, and biomechanics in need of realistic digital human representation.

\begin{table*}
  \caption{Comparison of \textit{GroundLink} with public MoCap and human pose estimation datasets.}
  \small
  \begin{tabular}{p{0.39\textwidth}cccccc}
    \toprule
    Dataset & $\#$ participants & $\#$ motions& $\#$ frames& $\#$ markers & Human body & Force type\\
    \midrule
    \citet{cmu} & 144 & 2435$^{*}$& 3.9M+$^{*}$ & 41 & Mesh$^\dag$ & N/A\\
    \citet{accad} &20$^{*}$&555$^{*}$& 192.6K+$^{*}$& 82 & Mesh$^\dag$ & N/A\\ 
    SFU \cite{sfu} &8$^{*}$&55$^{*}$& 109.7K+$^{*}$& 53 & Mesh$^\dag$ & N/A\\
    TotalCapture \cite{totalcapture2017} &5&60& 1.89M& 53 &Mesh$^\dag$& N/A\\
    Transitions \cite{AMASS:ICCV:2019}&1&110&108.7K  & 53 & Mesh & N/A \\ 
    Taiji Stability (PSU-TMM100) \cite{taichiScott2020}& 10$^{**}$&100$^{**}$& 1.36M& N/A &Skeleton& Pressure (vGRF)\\
    
    UnderPressure \cite{mourot2022underpressure} &10&218$^\ddag$&2.02M$^\ddag$ & 17$^\mathsection$ & Skeleton & Pressure (vGRF)\\
    
    \textbf{GroundLink (ours)} &\textbf{7}&\textbf{368}& \textbf{1.59M}& \textbf{96} & \textbf{Mesh} & \textbf{GRF, CoP}\\
  \bottomrule

\end{tabular}
\raggedright 
\footnotesize
\newline
$^*$ Online resource (accessed: May 2023). \newline
$^\dag$ Body surface mesh provided via processing by AMASS \cite{AMASS:ICCV:2019}. \newline
$^{**}$ Number of participants and motion sequence obtained from \citet{taichiScott2020} and \citet{collins2023lightweight}. \newline
$^\ddag$ Number of motions given by dataset downloaded from public resource. Number of frames calculated based on 5.6 hours of data at 100 fps provided in \citet{mourot2022underpressure}. \newline
$^\mathsection$ Motions tracked via 17 inertial measurement units (IMU). 
\label{table:mocap}
\end{table*}

\section{Related Work}


\subsection{Motion Capture with Contact Dynamics} 
\subsubsection{MoCap Dataset with Contact}
Several publicly-available human motion datasets are used in the graphics community (e.g. \cite{AMASS:ICCV:2019, cmu, totalcaptureTrumble:BMVC:2017, sfu, accad}). However, motion capture datasets that contain ground truth contact information are scarce, especially datasets with continuous foot contact. \citet{mourot2022underpressure} and \citet{taichiScott2020} released valuable datasets of motion capture with synchronized pressure data that is recorded via pressure insoles and include deep neural networks to estimate pressure distributions from 3D joint positions. \citet{Yin2003FootSeeAI} also designed an intuitive interface for digital avatar animation with their collected foot pressure sensor data. The datasets and tools serve specifically for animation usage and human stability analysis that contains locomotion, 24-form simplified Taichi movements and weight transfering tasks. Further, the dataset released from \citet{intelligentCarpetMIT} enables pose estimation from tactile signals given by pressure maps. However, pressure values captured with insoles or other types of pressure pads only contain contact information in magnitude (vertical GRF).  Additionally, the computer vision community has invested effort in gathering video-based datasets focused on various body parts (e.g. hand pressure estimation from sensor data from \citet{grady2022pressurevision} ) and scene interactions to address contact issues.

On the other hand, GRF and CoP are commonly understood concepts in biomechanics studies. Many motion datasets and experimental data that include kinematics, either full-body or focused on lower-extremity, along with the paired GRFs that are captured from ground force platforms or instrumented treadmill are available in the biomechanics community, which were released for gait analysis (e.g. \cite{4gait, shahabpoor2017measurement, walk1, walk2, walk3}), balance (e.g. \cite{balanceData1, balanceData2, WANG2020108580, Wang2020StandingBE}), or running (e.g. \cite{runningGRFmetrics, running2, HAMNER2013780}). These datasets often focus on a single specific motion type for patients' recovery and healthcare purposes. Further, there are existing datasets and efforts in incorporating contact labeling tasks such as \citet{shimada2022hulc}, where it provides scene geometry aware motion with estimated contacts guidance. We believe that including such contact labeling information in the current setup will eliminate potential grounding error introduced in the present method. Our collected dataset is unique in a way that it not only incorporates high-accuracy force data for GRF and CoP, but it also covers a variety of motion types, including subtle and balancing movements.

\subsubsection{Contact Dynamics Applications}
The measured data has proven to be highly valuable for foot-ground contact detection, such as the human foot keypoint dataset from \citet{cao2018openpose} for discrete contact labeling. Such foot contacts and contact detection have been extensively considered in character motion (e.g. \cite{10.1007/3-540-49384-0_6, 10.1145/566654.566607} ) in order to avoid and eliminate foot artifacts to produce contact-rich character control for more realistic virtual human motion and interactions (\cite{dlmotionsynthesisHolden16, phasennHolden17, he2018, starke2019}). More than that, physics-based simulation also often incorporates contact information with contact forces modeled in the physics engines (\cite{2019-MIG-symmetry, 10.1145/3386569.3392432, physicsWon2019}). Therefore, by providing more data grounded in real-world physics, such as ground reaction force, we aim to advance the realism of physical interactions of human movements with the environment.  

\subsection{Ground Reaction Force and Center of Pressure Estimation} Measurement of ground reaction forces and moments (GRF\&M) and Center of Pressure (CoP) is common in clinical biomechanics applications. Typically, these quantities are measured using force platforms that have a high spatial and temporal resolution. In sports science, the magnitude and rates of GRFs have been used to examine the risk of injuries in runners and other athletes (\cite{vanderWorp2016, DeBleecker2020}). In physical rehabilitation, GRFs often provide important information about disease mechanisms and can serve as therapeutic targets (\cite{Awad2020,COSTELLO20211138}). It is important to note that the horizontal components of GRF (i.e., anterior-posterior or medial-lateral) were found to be more important than the vertical component (vGRF) in these prior studies in people with stroke or those with knee pain.  In older adults, features of CoP displacement can predict the risk of falls  and are used as outcomes in fall prevention research (\cite{Quijoux2020}).

Measurement of GRFs and CoP with high accuracy requires laboratory-based methods with 3D motion and force capture systems. This can be expensive and restrictive ~\cite{estimateKAM}. Therefore, estimation of these quantities from other data (e.g., kinematics, inertial data) has been of interest in the biomechanics community with both machine learning (\cite{estvgrf, estGRFjump, estGRFml, predGRFML, MUNDT202029, s23010078}) and physics-based approaches (~\cite{estGRFspring, skals, VERHEUL2019716, BOBBERT19911095, massspring}) used in the past. Also,  portable solutions (e.g. wearable sensors \cite{POGSON202082}) using pressure insoles, typically restricted to vertical GRF, (e.g. \cite{mourot2022underpressure, taichiScott2020}) or inertial sensors (e.g. \cite{IMUs17010075, estvgrf, 9130158}) for estimating GRFs outside the lab in people's natural environments have been attempted. 

In the field of computer vision, researchers approach this problem to estimate the contact from monocular images and videos. With video-based motion and pose estimation, physics including contact is solved via optimization to create physics-aware motion (e.g. \cite{zell2017, li2019motionforcesfromvideo, physcapShimada2020, PhysAwareTOG2021}). \citet{li2019motionforcesfromvideo} proposed 3D motion and force estimation from videos through optimizing discrepancy between the observed and reprojected 2D joint and object endpoint positions. \citet{estimatecontactdynamics5459407} and \citet{physcapShimada2020} proposed video-based tracking for solving contact and performing physics-based contact labeling to prevent floor penetration for dynamic motions. \textit{GRFNet} proposed by \citet{PhysAwareTOG2021} estimates the GRF and compared it with the measured force provided by \citet{shahabpoor2017measurement} for gait movements. For broader body contact beyond foot-ground, \citet{clever2021bodypressure} inferred full-body structure and contact pressure using depth images related to body-mattress contact and \citet{Huang:CVPR:2022} utilized RICH which includes vertex-level contact labels on the body to reconstruct 3D contact-aware human body from 2D human scene interaction.

So far, 3D GRF and CoP measurements are primarily available within the biomechanics community. The closest related datasets, \citet{mourot2022underpressure} and \citet{taichiScott2020}, contain contact measurements, specifically pressure data which only denotes the vertical component of the ground reaction force.  Relatedly, prediction models such as \textit{GRFNet} from \citet{PhysAwareTOG2021}  provide evaluations with ground truth measurement for gait motion. Our dataset, \textit{GroundLink} contains full-body kinematics with 3D GRF and 2D CoP measurements for a variety of human movements with a particular emphasis on subtle motions.  

%% file: SecMocap.tex
\section{Data Acquisition}

\subsection{Overview}

\textit{GroundLink} is a MoCap dataset where each motion trial has full-body kinematics with paired GRF and CoP with annotation to distinguish between left and right feet. The force data were captured at high temporal resolution using force platforms that are embedded in the floor and have high measurement accuracy. We aim to use these data in computer graphics applications to create highly accurate animation and depict the movement and interaction of humans with the environment. 

We provide two different representations for human motion in our dataset as described in Sec.~\ref{sec:dataprocess}. The motion data is accompanied by the measured force data for the GRF and the CoP. The full pipeline of data acquisition and post-processing is illustrated in Fig.~\ref{fig:pipeline}. Compared to other available datasets for MoCap and contact (Table.~\ref{table:mocap}), such additional annotation for forces provides more comprehensive contact information for the human body, specifically for the foot segments.

\subsection{Data Capture}
This section describes the acquisition protocol for motion capture, GRF, and CoP data in a biomechanics lab as shown in Figs.~\ref{fig:teaser}, \ref{fig:pipeline}.

\subsubsection{Participants and Motions} We invited 7 participants (4 female and 3 male, age range 18–32) and their body dimensions are shown in Table~\ref{table:participant}. Participants performed 19 movements shown in Table~\ref{table:motion} with a focus on subtle motions. All stationary poses were held for approximately 20 $sec$, and most of the motions were performed with the participant making contact with the force plates. For the locomotion activities (e.g. walking),  it was ensured that the acceleration and deceleration when starting and ending the movement occurred outside of the force plates. Data from the three steps that included contact with the force plates were retained for these motions.  To ensure accurate measurement of GRF and CoP for both feet, each foot needs to make contact with a separate force plate during capture. Across all trials, data where only one body segment was in contact with the force plate at any instance were retained for further analyses.  Annotation for left and right foot was performed during post processing given the correspondence of the force plate ID and grounded foot. Participants performed four trials of each motion and data from three trials with highest quality were processed and included in the dataset.

\subsubsection{Motion Capture} To measure both the kinematics of the joints and the shape of the human body, 96 retroreflective markers (14 mm diameter) were placed on each participant. The full list of markers is illustrated in Fig.~\ref{fig:markerset} and we have provided detailed marker locations in the supplement. Marker data were captured with an infrared optical motion capture system (Qualisys, Goteberg, Sweden) ~\cite{qualisys} with 20 cameras at a sampling frequency of 250 $Hz$, A single RGB camera was used for video reference.

\begin{figure}[t]
\centering
\setlength{\unitlength}{\linewidth}
  \begin{picture}(1,0.55)%
      \put(0,0){\includegraphics[width=\linewidth]{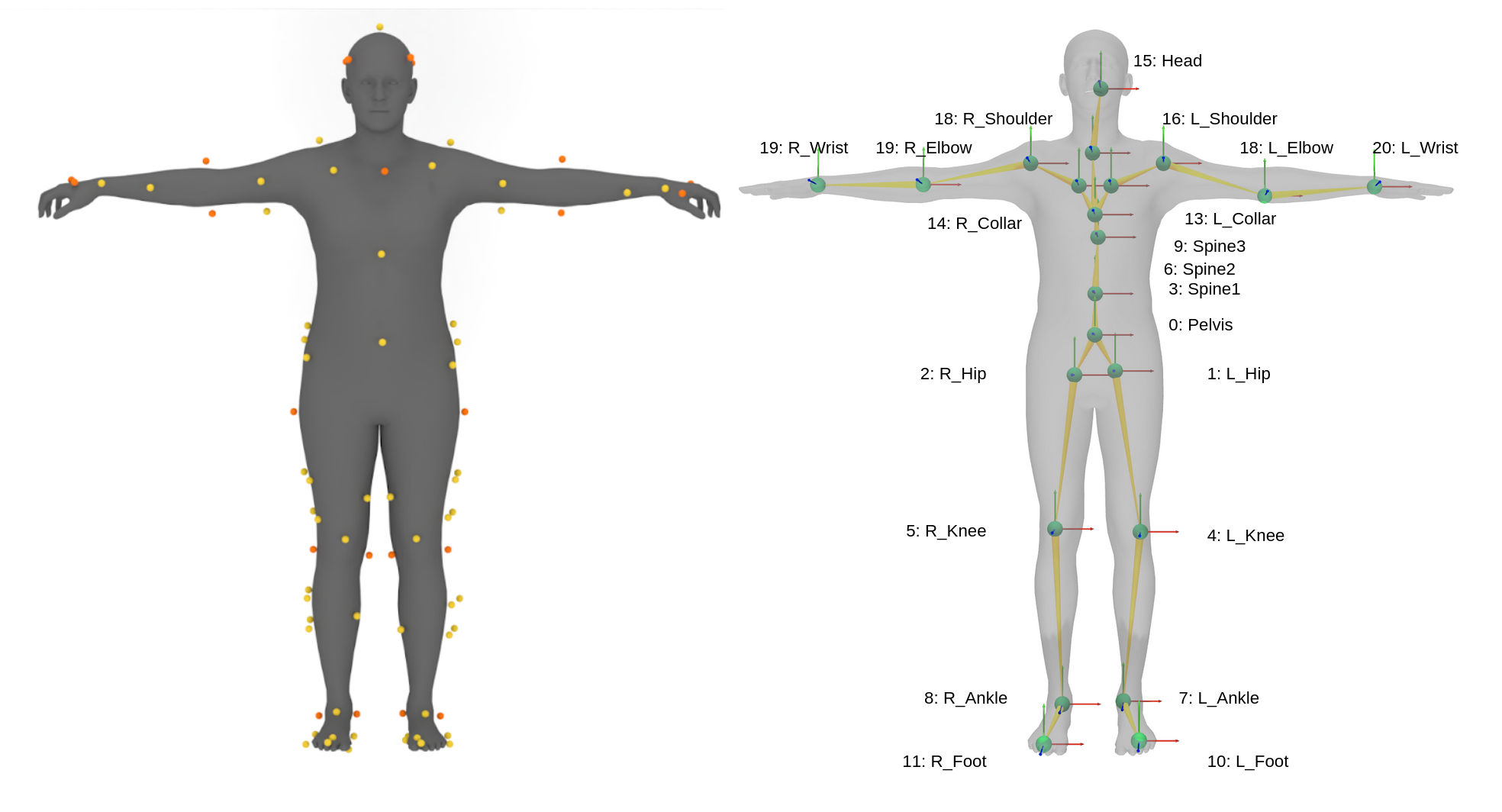}}%

    \small
  \end{picture}
\vspace{-20pt}
\caption{
Marker and skeleton setup. (Left) We use a custom placement with 96 markers in total. Our setup focuses on foot measurements with 10 markers on each foot. Markers used to solve pose and shape components in MoSh++ are colored orange (26 total). (Right) Kinematics tree extracted from MoSh++ in the order of input pose parameters ~\cite{SMPL-X:2019} used in \textit{GroundLinkNet}.
}
\label{fig:markerset}
\end{figure}

\subsubsection{Ground Reaction Force and Moments} 
\begin{figure}[t]
\centering
\setlength{\unitlength}{\linewidth}
  \begin{picture}(1,0.48)%
      \put(0,0){\includegraphics[width=\linewidth]{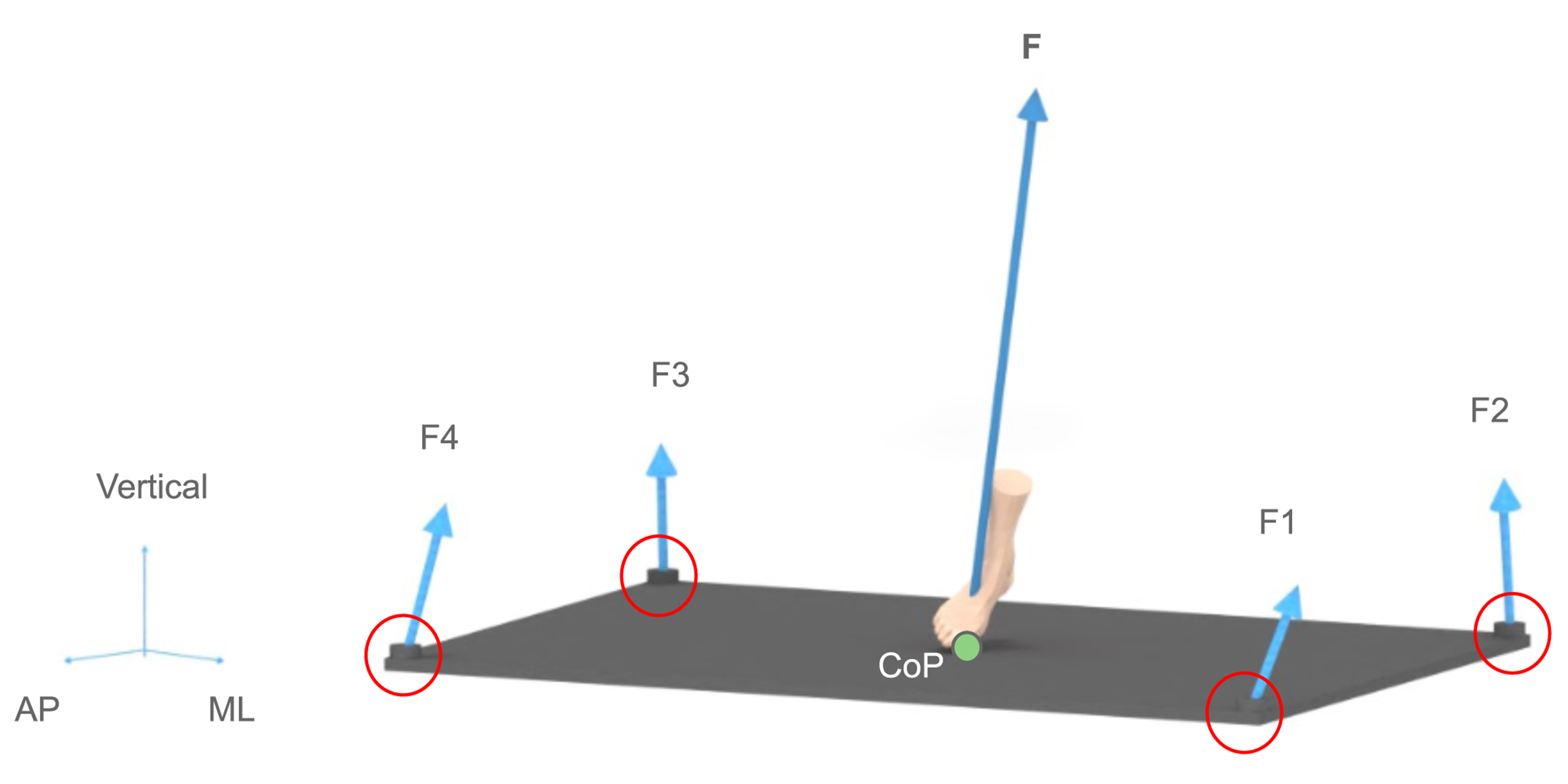}}%
    \small
  \end{picture}
\vspace{-20pt}
\caption{
Force plate illustration. Each force plate has four tri-axial force sensors (red circled) embedded in the plate that measures the force between the foot and the ground in the 3 axes: x (anteroposterior, AP), y (mediolateral, ML), and z (vertical) given the illustrated body position. The vectors $\mathbf{F}_i$ show the reaction forces measured by the force sensors. The sum of the measured forces is equivalent to GRF, $\mathbf{F}$. CoP is the point of application of $\mathbf{F}$. 
}
\label{fig:forceplate}

\end{figure}

Force plates, as shown in Fig.~\ref{fig:forceplate}, are specialized instruments embedded in the floor to measure the ground reaction force and center of pressure during movements. We used a total of five force plates with a sampling frequency of 2000 $Hz$. The GRF data were low-pass filtered at 20 Hz during post-processing. The GRF denoted as $\mathbf{F}$ refers to the reaction to the force that the body exerts on the ground upon contact, which has three orthogonal components that are typically denoted as $F_x$, $F_y$ and $F_z$ along the axes of the global coordinate system. The correspondence of these components with the anteroposterior (AP), mediolateral (ML), and vertical axes of the body can vary by convention used in the lab and orientation of the body relative to the global coordinate system. The sum of the measured forces is equivalent to the GRF, $\mathbf{F} = \sum_{i=1}^4\mathbf{F}_i$. Further, ground reaction moments or torques, denoted as $\mathbf{M}$, which represent the torques exerted by the GRF about the anterior-posterior, medial-lateral, and vertical axes, are also captured with the force plates.  

\label{sec:forcecapture}

\subsubsection{Center of Pressure} The center of pressure is the mathematical representation of the point of application of the ground reaction force, acting on the force plate. CoPs are calculated with the ground reaction forces and moments (GRF \& M) obtained from the force plates. By the definition, let $\mathrm{CoP} = (\mathrm{CoP}_x, \mathrm{CoP}_y, 0)$, which is a point always on the ground plane. Recall $\mathbf{F}$ and $\mathbf{M}$ are the ground reaction force and moment, respectively. The CoP coordinates can be calculated as:

\begin{equation} \label{equation:components}
\begin{split}
    \mathrm{CoP}_x = -\frac{M_y}{F_z} \;\;\;\; \mathrm{CoP}_y &= \frac{M_x}{F_z}
\end{split}
\end{equation}
\label{sec:cop}

\subsection{Data Post Processing} 
\label{sec:dataprocess}
We reconstruct the human body through different methods and representations. First, we label the raw data for all 96 markers, then we use both inverse kinematic solver embedded in Qualisys Track Manager (QTM) ~\cite{qualisys} and optimization proposed from Mosh++ ~\cite{AMASS:ICCV:2019} to solve kinematics and additional shape representations for surface reconstruction. During the process, different subsets of the markers are used, and both representations of the data will be released. 
\subsubsection{Labeling and Skeleton Construction}
We annotated the raw MoCap data in QTM to estimate the marker trajectories. Fully labeled data in c3d format with all 96 markers will be provided in the dataset. We also reconstruct the skeleton from 41 markers at joint positions for all the participants, with joint positions $\mathbf{p}\in \mathbb{R}^{23 \times 3}$ for all 23 joints in the Euclidean space. This skeleton tracking step is processed with QTM with a skeleton solver and skeleton template generated from the markers mapping. This hierarchical structure can be easily compatible and supported by a variety of 3D applications. Note that this representation was only used for comparative analysis with previous work~\cite{mourot2022underpressure} after skeleton retargeting, it will be released but was not included in the training and evaluation for GRF and CoP prediction because of the reasons listed in Sec ~\ref{sec:inputdata}.

\subsubsection{Joint and Surface Construction} From the labeled MoCap data, we adopted MoSh++ ~\cite{AMASS:ICCV:2019} to fit SMPL-X model ~\cite{SMPL-X:2019} and estimate both pose parameters, $\theta$ and shape parameters $\beta$ to reconstruct the human body. From the 96 markers, a subset of 26 markers is selected and used to reconstruct the human body surface shape and estimate the joint parameters, which are carefully selected given by the optimized markers introduced in ~\cite{AMASS:ICCV:2019} for the best reconstruction results. An initial guess is given for the markers and vertices' distances to the body surface, and we use MoSh++ ~\cite{AMASS:ICCV:2019} to refine the correspondence. Then, we optimize the parameters with a template from SMPL-X ~\cite{SMPL-X:2019} to minimize the distance between the markers and the corresponding vertices on the human body mesh. Details of the subset of the markers are shown in Fig. ~\ref{fig:markerset}.

\begin{table}
  \caption{Description of participants. We introduce diverse body shapes and sizes in \textit{GroundLink}.}
  \small
  \begin{tabular}{lccccccc}
    \toprule
    Participant ID & s001 & s002 & s003 & s004 & s005 & s006 & s007 \\
    \midrule
    Gender$^\dag$ & F & M & F & M & M & F& F\\
    Height (m) & 1.68 & 1.70 & 1.59 & 1.82 & 1.95 & 1.53 & 1.75\\
    Mass (kg) & 69.86 & 66.68 & 53.07 & 71.67 & 90.7 & 48.99 & 63.96\\
  \bottomrule
 
\end{tabular}
\raggedright
\footnotesize
$^\dag$ F: Female; M: Male.
\label{table:participant}
\end{table}

\begin{table}
  \caption{Description of motions. We focus on subtle movements and weight-shifting effects of motions. During the capture, each stationary pose is maintained for at least 20 sec to capture the balancing of the participant.}
  \small
  \begin{tabular}{lcc}
    \toprule
    Motion & Type & Duration$^\dag$ (sec)\\
    \midrule
    Tree (arms down) & yoga & 24.70\\
    Tree (arms up) & yoga & 24.40\\
    Chair & yoga & 23.98\\
    Warrior I & yoga & 24.34\\
    Warrior II & yoga & 25.03\\
    Dog & yoga & 19.85\\
    Side Stretch & yoga & 25.17\\
    Lambada dance & dynamic: dancing & 21.48\\
    High leg & dynamic: ballet & 13.87\\
    Small jump & dynamic: ballet & 13.43\\
    Squat & dynamic: exercise & 12.46\\
    Jumping Jack & dynamic: exercise & 13.95\\
    Stationary hopping & dynamic: exercise & 12.14 \\
    Taichi & dynamic: exercise & 14.81\\
    Swing & sports: tennis & 6.68\\
    Serving & sports: tennis & 7.66\\
    Kicking & sports: soccer & 4.81\\
    Walking & locomotion & 4.18\\
    Casual stand & idling & 16.77\\
  \bottomrule
  \footnotesize
  $^\dag$ Avg across all participants 
 
\end{tabular}

\label{table:motion}
\end{table}

%% file: SecNN.tex
\section{Experiments}

\subsection{Benchmark Neural Network Model}
\begin{figure}[t]
\centering
\setlength{\unitlength}{\linewidth}
  \begin{picture}(1,0.5)%
      \put(0,0){\includegraphics[width=\linewidth]{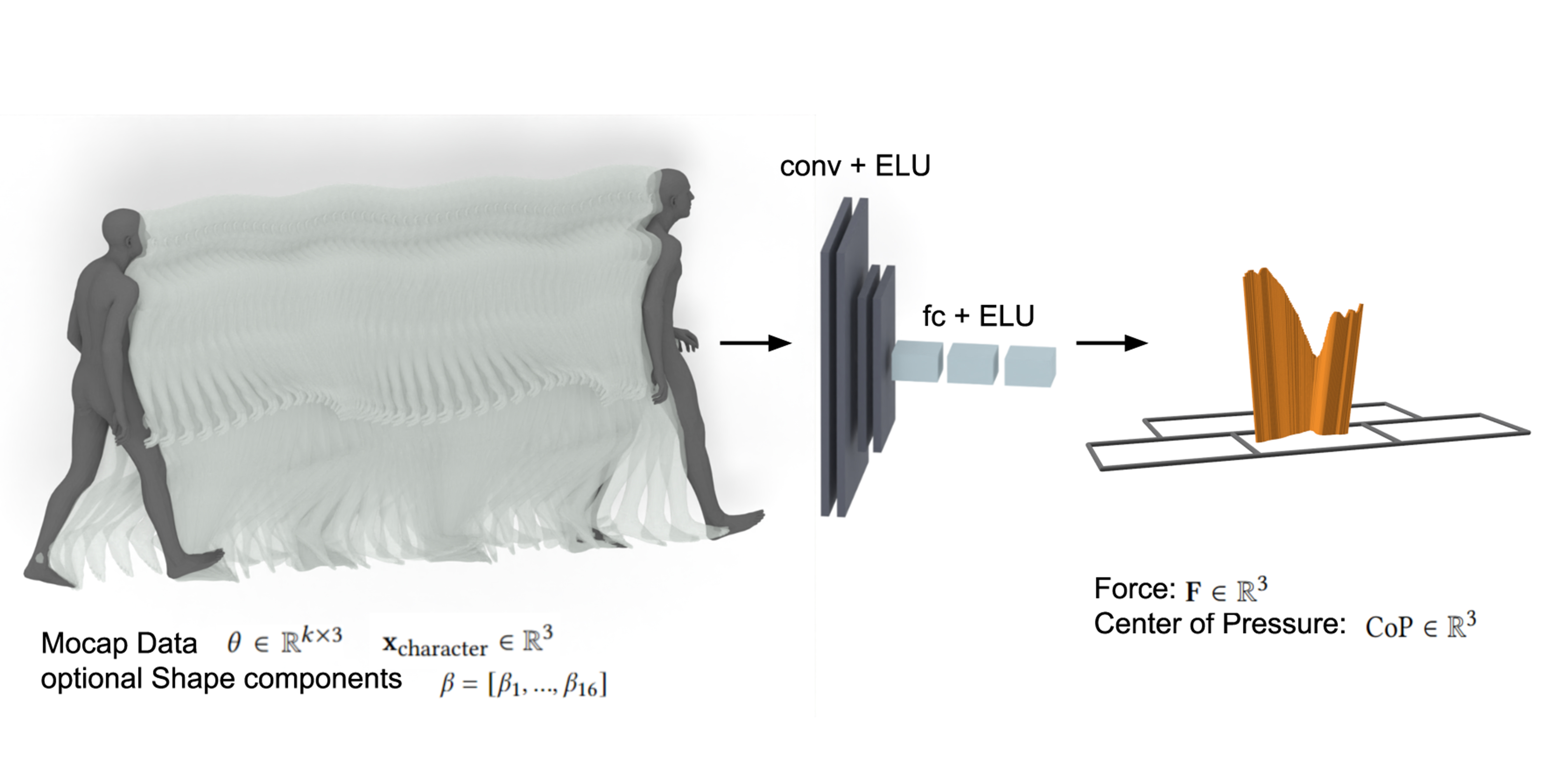}}%

    \small
  \end{picture}
\vspace{-20pt}
\caption{
Overview of \textit{GroundLinkNet}, our benchmark neural network model trained with \textit{GroundLink} to predict the GRFs and CoPs given the SMPL~\cite{SMPL-X:2019} pose and shape parameters.
}
\label{fig:nn}
\end{figure}

We prepared \textit{GroundLinkNet} (Fig.~\ref{fig:nn}), a deep neural network with a straightforward architecture trained with \textit{GroundLink}, to validate the effectiveness of having a dataset with kinematics paired by physics quantities.

\paragraph{Data representation}
\label{sec:inputdata} 
GroundLinkNet primarily takes relative joint angles $\mathbf{\theta} \in \mathbb{R}^{k \times 3}$ for $k$ joints and pelvis trajectories $\mathbf{x}_\mathrm{character}\in \mathbb{R}^3$ as inputs, with optional body shape components $\mathbf{\beta} = [\beta_1, ..., \beta_{16}]$, and will output the predicted GRF $\mathbf{F}\in \mathbb{R}^3$ and $\mathrm{CoP}_{\mathrm{character}}\in \mathbb{R}^3$.

Using global coordinates as described in ~\cite{Mourot22} is not ideal due to its lack of invariance of translation and rotation with respect to the ground plane. Stochastic data augmentation as proposed in ~\cite{Mourot22} cannot generalize to the arbitrary placement of characters in the scene while increasing the training time. 
Hence, we represent the data with relative positions and joint orientation to avoid confusion caused by using global representation in the coordinate system. Further, due to the double-cover of raw quaternions ~\cite{quaternion10.1007/s11263-019-01245-6}, we follow a similar setup as ~\cite{supertrack10.1145/3478513.3480527} to introduce our data representation.

First, we adjust the joint rotation representation. Instead of using global position, $\mathbf{x}_{k_i}$, the position for $k$th joint at frame $i$, we use the local rotation represented by rotation vectors of each joint with respect to the parent joint in the kinematics tree, as shown in Fig.~\ref{fig:markerset}, $\mathbf{\theta} \in \mathbb{R}^{k \times 3}$. Then, we apply transformations to both the center of pressure and the pelvis position to avoid using any global representation in the network and transform to the character space. Given the definition of the center of pressure, since it is always located on the ground surface with $\mathrm{CoP}_{\mathrm{global}} = (\mathrm{CoP}_x, \mathrm{CoP}_y, 0)$ for both left and right foot, we use the projected pelvis as the origin for the character space. Let $T$ be the transformation matrix of the pelvis projected on the ground plane, which is $xy-$plane with z-positive system, and consists of the rotation and translation of the pelvis. Then, the relative positions of the pelvis and the center of pressure are given as: 

\begin{equation}
    \mathbf{x}_\mathrm{character} = T^{-1} \cdot  \mathbf{x}_\mathrm{pelvis} = (0,0,\mathbf{x}_z)
\end{equation}

\begin{equation}
    \mathrm{CoP}_{\mathrm{character}} = T^{-1} \cdot  \mathrm{CoP}_{\mathrm{global}}
\end{equation}
Lastly, shape parameters with 16 components solved by MoSh++ ~\cite{AMASS:ICCV:2019} denoted as $\mathbf{\beta} = [\beta_1, ..., \beta_{16}]$ defined by PCA coefficients in shape space are additional input features. In the end, the input features are the concatenation of the pose $\mathbf{\theta}$, $\mathbf{x}_\mathrm{character}$, and the optional shape $\mathbf{\beta}$. The output features are the concatenation of the GRF $\mathbf{F}$ and $\mathrm{CoP}_{\mathrm{character}}$.

\paragraph{Network Architecture} ~\citet{Mourot22} proposed and structured a nerual network to accommodate motion sequences with variable lengths. The architecture of the network comprises four 1D temporal convolutional layers, each layer with kernels spanning 7 frames. These are followed by three fully connected layers, each independently applied to every frame to maintain adaptability with variable-length sequences. After each convolutional or fully connected layer, exponential linear units (ELU) serve as the activation function, with the exception of the final layer where a softplus activation function is deployed to ensure that the relative pressure values of the sensors from different regions, namely the vertical ground reaction force components, are outputted nonnegative. In our experiments, on the other hand, it is important to note that the components that are predicted - namely the ground reaction force (excluding the vertical GRF) and the center of pressure --- need not be restricted to positive values. Both the center of pressure and ground reaction force can possess negative values, reflecting the directional attributes of these vectors.

\paragraph{Training}

Since both direction and magnitude are significant to reconstruct GRF and CoP, we use mean squared error (MSE) to measure the difference between the predicted and target values. The loss function is then defined as:

\begin{equation}
    \mathcal{L} = \frac{1}{N} \sum_{i=1}^N \|\mathbf{F}_i - \mathbf{\hat{F}}_i\|^2
\end{equation}
where $\mathbf{\hat{F}}$ is the predicted GRF by the neural network. 
 
For evaluation, we use s007's data as a testing set. When performing cross-validation, a different participant serves as a testing subject. With the remaining served as training data, we split it with a ratio of $7:3$ for the training and validation set.

\begin{figure}[t]
\centering
\setlength{\unitlength}{\linewidth}
  \begin{picture}(1,0.5)%
      \put(0,0){\includegraphics[width=\linewidth]{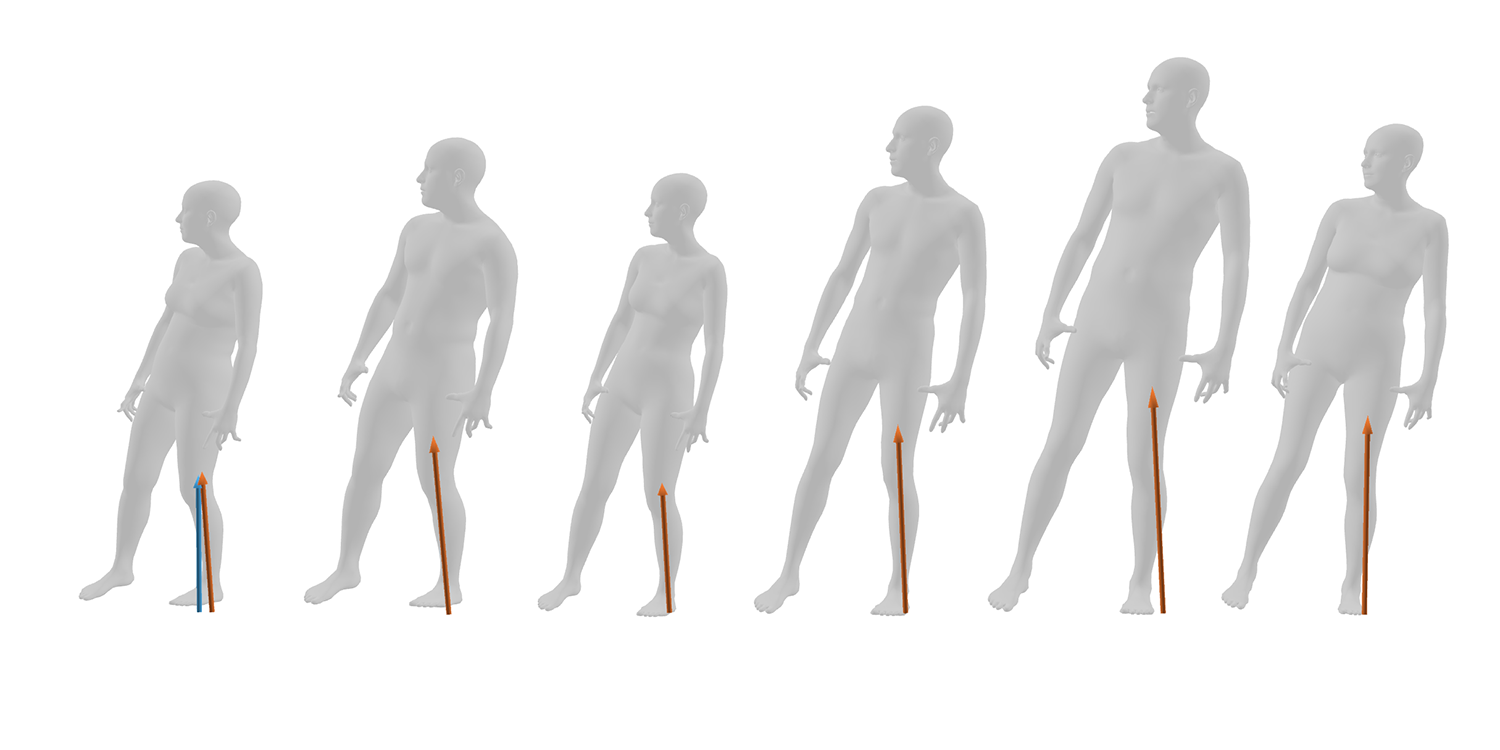}}%

    \small
    \put(0.035,0.40){(a)}%
    \put(0.2,0.40){(b)}%
  \end{picture}
\vspace{-20pt}
\caption{
Shape-aware force prediction. Given the same pose from (a) and various body shapes in (b), our \textit{GroundLinkNet} can predict plausible GRFs respecting different body sizes. 
}
\label{fig:shape}
\end{figure}

%% file: SecResult.tex
\subsection{Evaluations}

\begin{figure}[t]
\centering
\setlength{\unitlength}{\linewidth}
  \begin{picture}(1,1.18)%
      \put(0,0){\includegraphics[width=\linewidth]{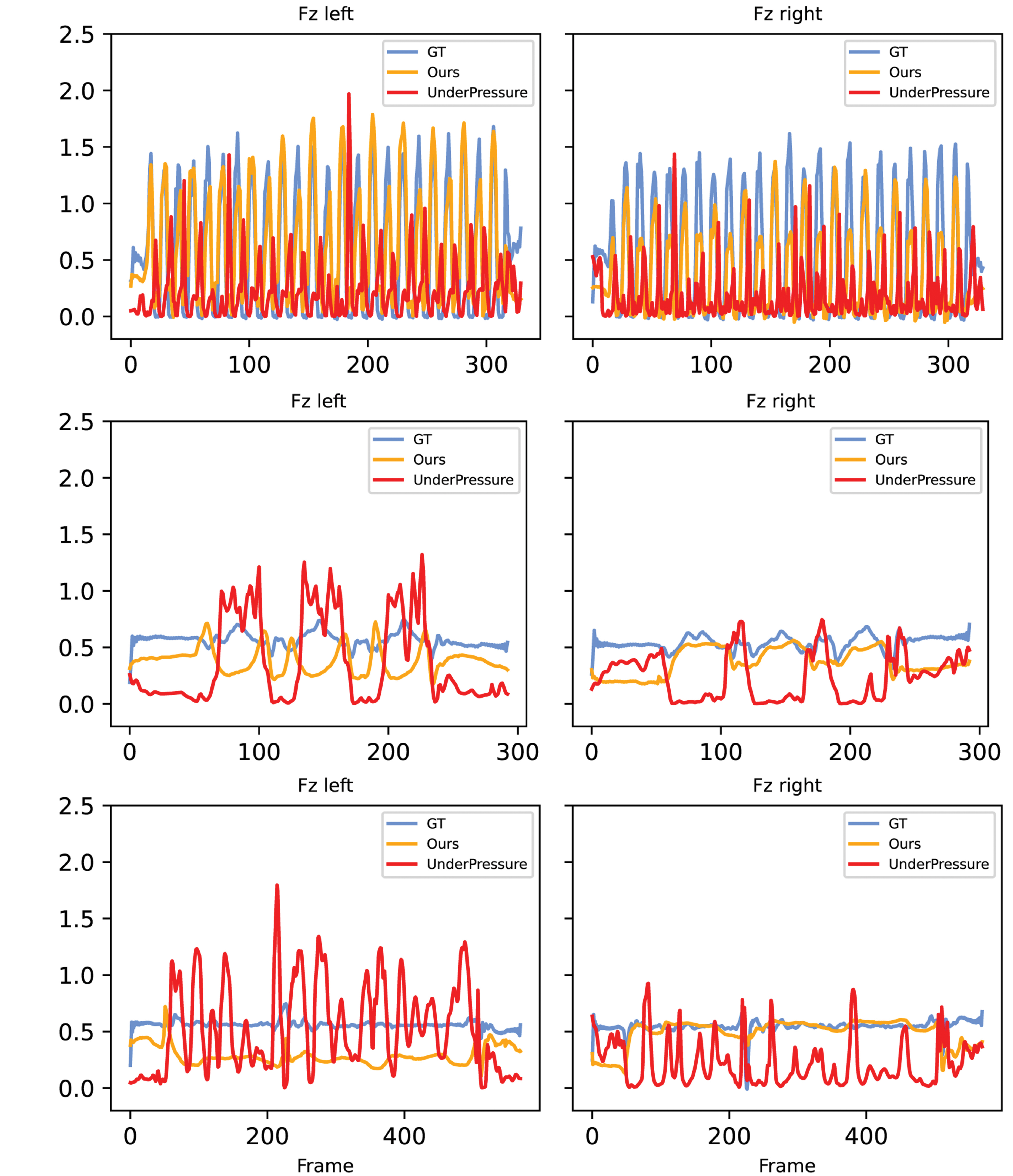}}%
    \small
    \put(0.004,1.121){(a)}%
    \put(0.004,0.738){(b)}%
    \put(0.004,0.36){(c)}%
  \end{picture}
\vspace{-20pt}
\caption{
Comparison against \textbf{U}nder\textbf{P}ressure ~\cite{mourot2022underpressure} for motions: (a) jumping jacks (b) squatting (c) chair. Horizontal axis: frame \# (downsampled by 1/10); vertical axis: force normalized by weight (test subject weighs 90.7kg). Our model yields improved predictions for the vertical component (vGRF), even though our model handles a more challenging task of predicting the full 3D GRF and CoP.
}\label{fig:upcomparison}

\end{figure}
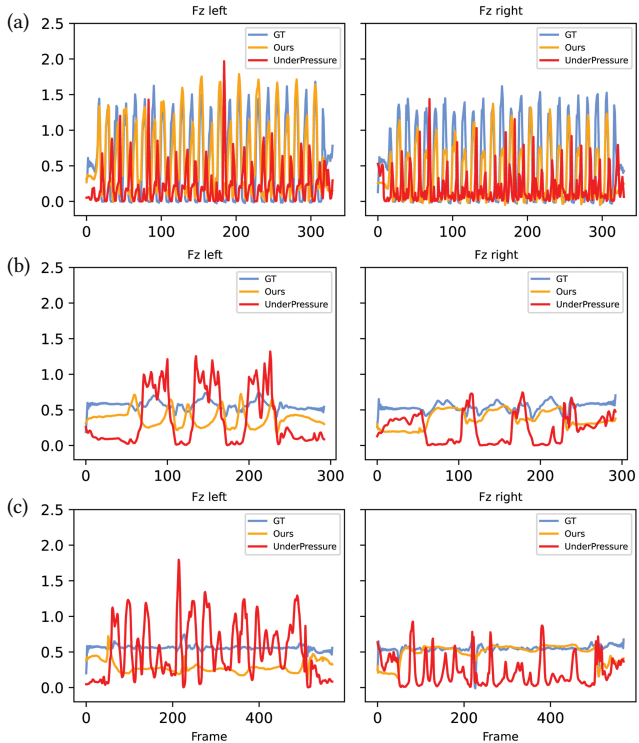

\begin{table}
  \caption{vGRF Prediction error comparison between \textit{GroundLinkNet} and \textbf{U}nder\textbf{P}ressure. Values are mean square error (MSE) with the target measured values normalized by weight (test subject weighs $90.7kg$).}
  \small
  \begin{tabular}{p{0.26\textwidth}cc}
    \toprule
    Motion & UnderPressure$^\dag$ & Ours$^\dag$\\
    \midrule
    Chair&0.94/0.19&0.08/0.02\\
    Tree (arm up)&0.54/0.11&0.49/0.03\\
    Tree (arm down)&0.78/0.10&0.14/0.03\\
    Warrior I&0.15/0.16&0.02/0.06\\
    Warrior II&0.09/0.09&0.02/0.04\\
    Dog&0.69/0.13&0.04/0.07\\
    Side Stretch&0.31/0.46&0.06/0.04\\
    Lambada dance&0.27/0.40&0.06/0.13\\
    Squat&0.16/0.14&0.06/0.04\\
    Jumping Jack&0.60/0.50&0.08/0.08\\
    Stationary Hopping&0.80/0.27&0.23/0.03\\
    Taichi&0.19/0.18&0.06/0.09\\
    Soccerkick&0.49/0.41&0.45/0.18\\
    Total&0.44/0.23&0.18/0.07\\
  \bottomrule
\end{tabular}
\raggedright
\footnotesize
$^\dag$ The error values are calculated errors for the left and right foot, normalized by body weight.
\label{table:upcomparison}
\end{table}

\label{sec:evaluation}
\paragraph{Testing with \textit{GroundLink}}
At the stage of inference, our model predicts the GRF and CoP components $F_x, F_y, F_z$ and $\mathrm{CoP}_x, \mathrm{CoP}_y, \mathrm{CoP}_z$, respectively, from the inputs of pelvis position, joint rotations, and shape components. To evaluate the robustness of the models, we perform cross-validation for all the participants (error report in Table~\ref{table:error}) and provide a report in Fig.~\ref{fig:s7}, which contains a comparison of ground truth and predicted GRF and CoP for three example motions: taichi, side stretching, and hopping movements. These examples not only demonstrate weight shifting between feet but also shows our model's performance on more dynamic movement such as jumping and stationary hopping. Further, we perform a comparative analysis with vGRF prediction provided by \textbf{U}nder\textbf{P}ressue ~\cite{mourot2022underpressure} and present the results in Fig.~\ref{fig:upcomparison} and Table~\ref{table:upcomparison}. 

\paragraph{Testing with kinematics-only data}
Many existing datasets have optimized pose and shape components through ~\cite{AMASS:ICCV:2019}, such as OSU ACCAD ~\cite{accad}. While they do not contain any force data associated, we can still evaluate the performance of the model visually. Among all the testing data that are available, it shows that our model predicts convincing GRFs for stationary poses such as martial arts punching and casual standing. We note that while our CoP predictions generally follow the foot positions, they can incorrectly offset from the feet, as observed by the arrows coming off from the feet in visualizations as shown in Fig.~\ref{fig:amass}(b). We believe this is due to the CoP representation relative to the pelvis. We show more results from different datasets in our supplementary video. 

\paragraph{Testing with various body shapes}
Body shape and body weight are factors that can affect the ground reaction force. Among all the components, vertical ground reaction force (vGRF) is the most sensitive to the body shape. With the shape parameters incorporated into the training setup, we evaluate the correspondence of the human body shape with both the ground reaction force and the center of pressure. The analysis of our predicted GRF for people with varying body shapes, as shown in Fig.~\ref{fig:shape}, reveals that individuals with larger physiques tend to exhibit higher vGRF ($F_z$) as expected.

\begin{table}
  \caption{Mean square error (MSE) with target measured values}
\small
  \begin{tabular}{p{0.105\textwidth}@{\hspace{0.1cm}}c@{\hspace{0.15cm}}c@{\hspace{0.15cm}}c@{\hspace{0.15cm}}c@{\hspace{0.15cm}}c}
    \toprule
    Motion & $F_x$ $^{\dag}$ & $F_y$ $^{\dag}$ & $F_z$ $^{\dag}$& $\mathrm{CoP}_x$ $^{\ddag}$ & $\mathrm{CoP}_y$ $^{\ddag}$\\
    \midrule
    Chair&0.12/0.13&1.85/1.59&25.62/20.54&1.46/6.56&2.61/3.54\\
    Tree (arm up) &0.05/0.03&1.34/0.45&149.8/14.55&2.09/8.72&1.56/2.37\\
    Tree (arm down) &0.05/0.02&1.30/0.41&130.8/17.08&3.88/14.68&0.70/3.28 \\
    Warrior I&0.13/0.14&1.96/1.78&22.80/66.49&1.19/31.90&4.63/14.83\\
    Warrior II&0.19/0.24&0.96/1.37&15.25/48.12&1.23/20.46&1.98/9.07\\
    Dog&0.65/0.76&1.80/2.54&55.34/86.38&12.1/10.82&13.46/48.52\\
    Side Stretch&0.22/0.26&3.22/4.09&43.36/51.04&9.26/49.63&4.26/20.39\\
    Lambada Dance&0.50/0.59&3.30/4.08&70.37/93.94&6.79/11.90&3.52/7.47\\
    Squat&0.14/0.15&1.27/1.30&24.36/28.07&1.71/4.27&1.71/2.49\\
    Jumping Jack&2.82/3.20&5.89/4.98&61.68/74.59&3.21/3.54&2.45/3.79\\
    Hopping$^{\ast}$&2.36/0.09&1.85/0.49&221.5/29.34&5.0/14.74&2.56/11.88\\
    Taichi &0.12/0.15&1.98/2.06&41.41/54.54&7.95/9.00&5.70/7.89\\
    Tennis: Swing&0.28/0.22&0.88/1.83&19.24/30.28&4.20/2.58&6.32/7.26\\
    Tennis: Serve&0.23/0.29&1.64/1.17&42.10/42.17&2.27/5.76&2.72/24.09\\
    Soccerkick&5.05/3.79&2.88/2.02&178.9/118.5&14.02/25.56&11.21/39.41\\
    Casual Stand&0.07/0.07&0.55/0.92&46.69/52.71&3.36/11.19&2.03/8.37\\
  \bottomrule

\end{tabular}
\raggedright
\footnotesize
$^{\ast}$ refers to Stationary Hopping.\newline
$^\dag$ The calculated error for left/right foot across all test subjects ($\times 10^{-3}$), given by force normalized by weight.\newline
$^\ddag$ Error for left/right foot across all test subject ($\times 10^{-3}$) in $m$
\label{table:error}
\end{table}

\section{Conclusion}
In this paper, we introduce \textit{GroundLink}, a new motion capture data with corresponding GRFs and CoPs, which are difficult to obtain due to the restricted resources of the lab and spaces. From the motion and contact force data, we present \textit{GroundLinkNet}, a benchmark model trained from \textit{GroundLink} to predict GRFs and CoPs from the kinematics and shapes of the human body. With the possibility of estimating the contact force from pure kinematics of human characters, we hope our data and model can serve numerous downstream applications of digital humans with physical realism.

\paragraph{Discussion and Future Work}

As a benchmark model, we noticed some prediction inconsistencies delivered by \textit{GroundLinkNet}. First, the model is currently set to use the projected pelvis as the character space origin, which results in a noticeable offset in predicting CoP, as illustrated in Fig.~\ref{fig:amass}(b). To remedy this, a more appropriate frame-of-reference, such as projected feet for referring left and right GRF and CoP, can enhance prediction accuracy. Second, \textit{GroundLink} only contains walking trials with a maximum of three-step strides to keep the participants on the limited region of the force plates. Because of this, we observe that our benchmark model struggles to predict GRFs for different types of locomotions.
Third, we also noticed the model being too sensitive to small changes in the pose for upper body, e.g. hand rotations, affecting the output. 
These are signs that there is not enough data (such as isolated upper body movements) to deal with such out-of-distribution test scenarios.
While we showed that \textit{GroundLinkNet} can generalize to unseen scenarios to some extent, it is also clear that the data scarcity issue must be addressed. We believe there are open research opportunities in (a) deriving a better machine learning scheme, including data representation, to understand physics from a limited amount of data, and (b) simulating physics to obtain more data, while also leveraging the captured physics quantities simultaneously.

%% file: SecFigsOnly.tex
\begin{figure*}[t]
\centering
\setlength{\unitlength}{\linewidth}
  \begin{picture}(1,0.22)%
      \put(0,0){\includegraphics[width=\linewidth]{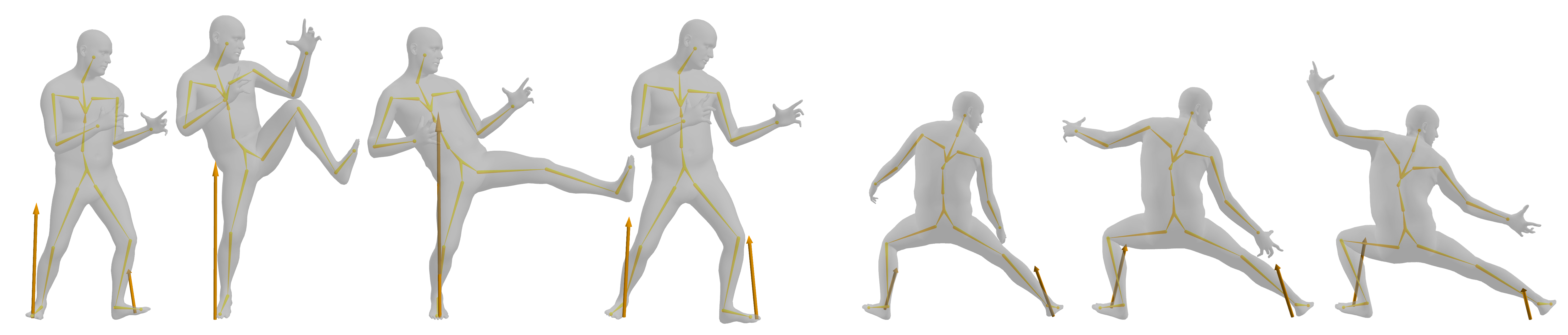}}%
    \small
    \put(0.01,0.2){(a)}%
    \put(0.55,0.2){(b)}%
  \end{picture}
\vspace{-20pt}
\caption{
Testing result for OSU ACCAD: (a) push kicking motion and (b) Victory martial arts. Preprocessed MoCap data is obtained from AMASS to retrieve the pose components for input parameters for the model. Even in the absence of ground truth contact forces data for external datasets, we employ our trained model from \textit{GroundLink} to predict GRF and CoP based solely on the available kinematics.}. 
\label{fig:amass}

\end{figure*}

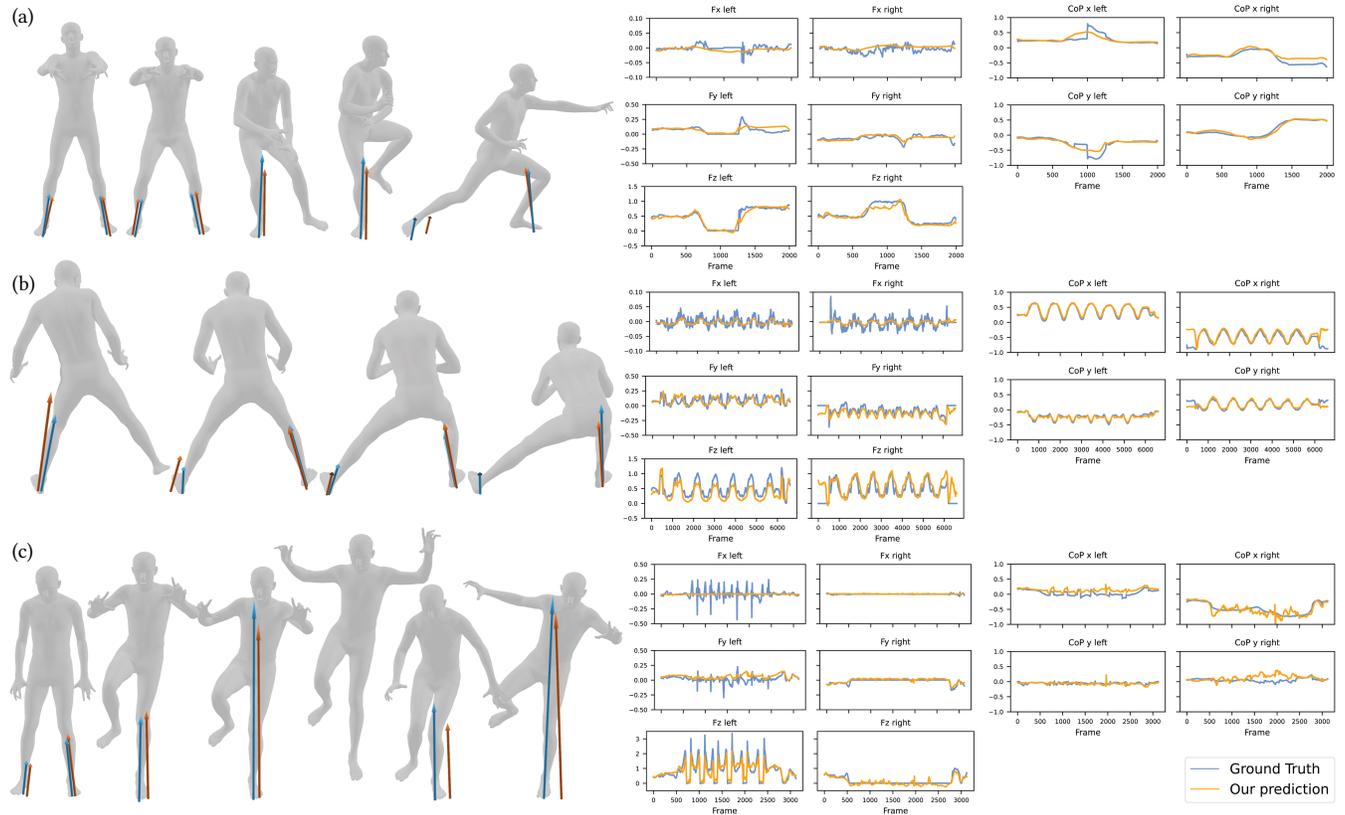
\begin{figure*}[t]
\centering
\setlength{\unitlength}{\linewidth}
  \begin{picture}(1,0.65)%
      \put(0,0){\includegraphics[width=\linewidth]{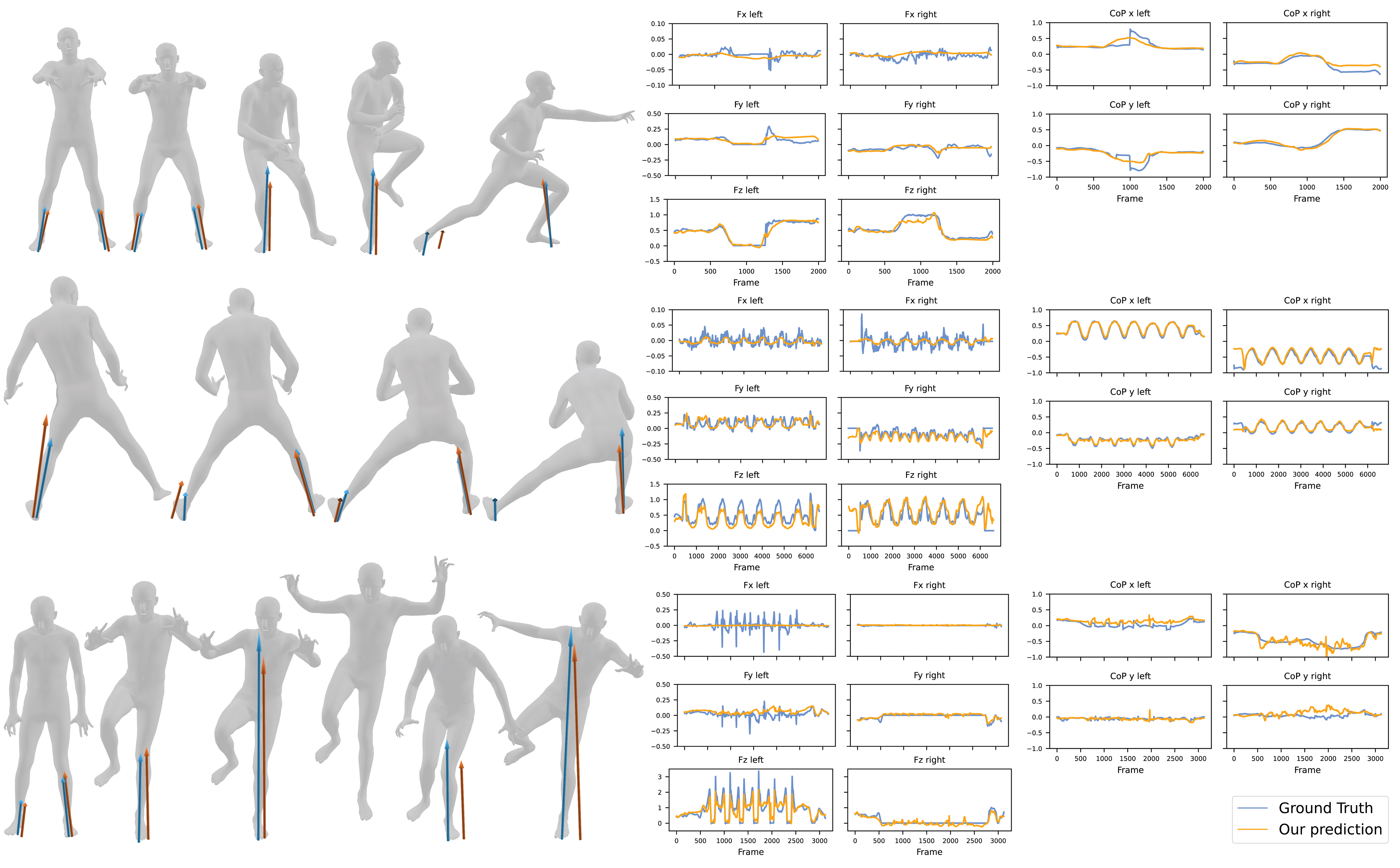}}%
    \small
    \put(0.004,0.6){(a)}%
    \put(0.004,0.4){(b)}%
    \put(0.004,0.2){(c)}%
  \end{picture}
\vspace{-20pt}
\caption{
\textit{GroundLinkNet} results for participant s004. The model is trained with the data from the other 6 participants. The plots show a comparison of the predicted value (orange) and the ground truth values (blue) for GRF (normalized by subject weight) and CoP ($m$). Horizontal axis: video frame \#; vertical axis (middle): force normalized by weight (subject weighs 71.67kg); vertical axis (right): relative position with projected pelvis ($m$).Example motions: (a) taichi, (b) side stretching, and (c) stationary hopping. Overall, the models generate reasonable prediction patterns for GRF for various motions. For CoP prediction, it performs better with smooth motions such as side stretching with consistent weight shifting patterns but is relatively less convincing for motions when a single foot is lifted and shifts dramatically in the air such as taichi movement.
}
\label{fig:s7}

\end{figure*}